\documentclass[global, twocolumn]{svjour}

\usepackage{amssymb, amsmath, amsfonts}
\usepackage{graphicx}
\usepackage{subfigure}
\usepackage{times}
\usepackage{cite}

\begin{document}


\title{Acoustic wave propagation in fluid metamaterial with solid inclusions} 



\author{I. V. Lisenkov\inst{1,2} \and R. S. Popov\inst{1} \and S. A. Nikitov\inst{1,2}}
\institute{Kotel'nikov Institute for Radio-engineering and Electronics of RAS \and Moscow Institute for Physics and Technology (State University)}
\mail{lisenkov@cplire.ru}

\begin{abstract}
Acoustic waves propagation of  in composite of water with embedded double-layered silicone resin/silver rods is considered. Approximate values of effective dynamical constitutive parameters are obtained. Frequency ranges of simultaneous negative constitutive parameters are found. Localized surface states on the interface between metamaterial and ``normal'' material are found. Doppler effect in metamaterial is considered. Presence of anomalous modes are shown.
\end{abstract}

\keywords{Metamaterials -- Bulk acoustic waves -- Surface acoustic waves -- Doppler Effect}

\maketitle 

\section{Introduction}
Media with effective negative constitutive parameters attract attention for several decades due to unique properties of wave propagation in them. In pioneering works of Veselago~\cite{Veselago68} and Pendry~\cite{Pendry} electromagnetic waves in media with simultaneous negative $\epsilon$ and $\mu$ were considered. Since that time, a lot of theoretical and experimental works has been done on this subject~\cite{Carloz05}. In fact, for acoustic waves metamaterials are possible too and for them both density and stiffness should be negative. Some proposals and experiments to achieve negative effective constitutive parameters for acoustic composites were made in the past, exploiting different techniques, such as: embedding soft inclusions in fluids~\cite{Liu2000,Jensen2004}, using Helmholtz resonators~\cite{Fang2006} and pipe-membrane structures~\cite{Lee2010}. 

The aim of this work is to show that it is possible to construct a metamaterial to exhibit negative constitutive parameters within ultrasonic range and demonstrate that acoustic waves can exhibit exotic behavior, namely presence of localized states at the boundary between metamaterial and ``normal'' material and reversed Doppler effect. In the mathematical model both longitudinal and shear modes are included in calculations. Also internal dissipation losses are taken in account.

\section{Mathematical model}
\subsection{Effective constitutive parameters}
In this work a composite consisting of fluid host with rods embedded in it is considered. Rods have round cylindrical shape and consist of elastic materials. Acoustic waves propagate in host material (density $\rho_0$ and bulk modulus $B_0$) with wave-vector perpendicular to generatixes of the rods. A metamaterial should be quasi-isotropic, so the wavelength is much longer than the distance between the centers of cylinders~($L$) and their radii~($R$)~\cite{Carloz05}. All cylinders are considered identical and are placed at random at approximate equal distance between neighbors. The aim is to compute the dynamical effective constitutive parameters of a metamaterial on given frequency and dispersion of the propagating bulk acoustic wave. 

To proceed the calculations coherent potential approximation (CPA)~\cite{Jensen2004,Sheng2006} is used. We used this method applied to fluid/solid composite taking into account longitudinal and shear polarizations in solid inclusions and acoustic dumping in both inclusions and host.

Following the CPA procedure, one inclusion is surrounded with cylinder of radius $L$ (forming coated inclusion) and outside the cylinder some effective fluid is placed. The constitutive parameters of effective fluid  ($\rho_e$ for effective density and $B_e$ for effective bulk modulus) are unknown and to should be estimated. Pressure ($\Psi(r,\phi,t)$) and radial component of velocity ($v_r(r,\phi,t)$) in host material and in effective material can be decomposed by cylindrical harmonics~\cite{Blackstock2000}:
\begin{equation}
\begin{aligned}
\Psi = e^{-i\omega t}\sum_{n=0}^\infty X^m_nJ_n(k_mr)e^{in\phi}+Y^m_nH_n(k_mr)e^{in\phi}\\
v_r = e^{-i\omega t}\dfrac{k_m}{\rho_m}\sum_{n=0}^\infty X^m_nJ'_n(k_mr)e^{in\phi}+Y^m_nH'_n(kr)e^{in\phi},
\end{aligned}
\label{eq:cyldecomp}
\end{equation}
where $J_n()$ and $H_n()$ are Bessel and Hankel functions of the first kind, $n$-th order, $X^m_n$, $Y^m_n$ are unknowns and $m$ corresponds to $e$ in effective media and for $0$ in host material.

On the interface between the host fluid and the effective medium standard boundary conditions of velocity and pressure continuity should be satisfied. Due to cylindrical symmetry and isotropy of materials, we could satisfy boundary conditions for each cylindrical harmonics independently~\cite{Blackstock2000}:
\begin{equation}
\begin{aligned}
 &\begin{bmatrix}J_n(k_eL)&H_n(k_eL)\\\dfrac{k_e}{\rho_e}J_n(k_eL)&\dfrac{k_e}{\rho_e}H_n(k_eL)\end{bmatrix}\cdot
\begin{bmatrix}X^e_n\\\vphantom{\dfrac{k_e}{\rho_e}}Y^e_n\end{bmatrix}= \\
 &\begin{bmatrix}J_n(k_0L)&H_n(k_0L)\\\dfrac{k_0}{\rho_0}J_n(k_0L)&\dfrac{k_0}{\rho_0}H_n(k_0L)\end{bmatrix}\cdot
\begin{bmatrix}X^0_n\\\vphantom{\dfrac{k_e}{\rho_e}}Y^0_n\end{bmatrix},
\end{aligned}
\label{eq:bc:coated} 
\end{equation}
where  $k_e$ is the wave-number in effective medium, $k_0$ is the wave-number in host medium
and $e^{in\phi}$ and $e^{-i\omega t}$ factors are dropped out.

Main CPA condition is an absence of scattering on the boundary, which means that there is a perfect matching between the coated inclusion and surrounding effective medium. Setting $X^e_n=1$ and $Y^e_n=0$ in~\eqref{eq:bc:coated} we obtain the CPA equation:
\begin{equation}
\begin{aligned}
 &-\dfrac{k_0}{B_0}J'_n(k_0L)J(k_eL) + \dfrac{k_e}{B_e}J_n(k_0L)J'(k_eL)=\\
\dfrac{Y_n}{X_n}&\left[\dfrac{k_0}{B_0}H'_n(k_0L)J(k_eL) - \dfrac{k_e}{B_e}H_n(k_0L)J'(k_eL)\right]
\label{eq:main:CPA}
\end{aligned}
\end{equation}
Exploiting the fact that metamaterial should be quasi-isotropic, in other words, the phase of the wave in host material should not change much on a  distance between neighbor inclusions ($k_0L\to0$ and $k_eL\to0$). We use asymptotic formulas for Bessel and Hankel functions for near-zero argument limit. For small argument approximation only two, namely, zero-th and first orders of Bessel and Hankel functions are significant. After some algebra we obtain approximate formulas for effective constitutive parameters:
\begin{align}
 B_e  = \dfrac{B_0}{1-S_0},&\,\,\,\,\, \rho_e = \rho_0 \dfrac{1-S_1}{1+S_1}
\label{eq:effective:const}
\end{align}
where:
\begin{equation}
 S_n=\dfrac{D_n}{1+D_n}\dfrac{4}{\pi}\dfrac{1}{k_0^2L^2}\cdot i
\end{equation}
and $D_n=Y_n/X_n$ is a scattering coefficient of the one embedded rod in an infinite fluid. In our case it is scattering of the acoustic wave in fluid on a round solid cylinder. This problem was widely investigated in past~\cite{Faran}.

As it is seen from formulas~\eqref{eq:effective:const}, dynamical effective constitutive parameters are frequency dependent and in the case of resonance inside inclusion their values can  be changed dramatically and even become negative. Thus the speed of acoustic wave inside the inclusions should be much smaller than in a host fluid, to form resonance frequency at which the wavelength in a host is much longer that distance between inclusions.

Also we should notice, that bulk modulus depends on a monopolar scattering coefficient and effective density depends on the dipole coefficient. This is true, because the monopole mode is the uniform compression or expansion of the inclusion and dipole mode is the inclusion shift.

The refractive index of the medium is described by the following formula:
\begin{equation}
 n(\omega)=\sqrt{1/B_e(\omega)}\cdot\sqrt{\rho_e(\omega)}
\label{fig:refind}
\end{equation}
If the argument of refractive index  tends to zero the wave-vector is pointing  same direction as the Poynting vector. Argument's value  close to  $\pi/2$ means that the one of the constitutive parameters is negative and the other positive, thus the wave finally decays. Argument's value close to  $\pi$ means that wave-vector and Poynting vector are directed oppositely and the wave have backward behavior.

\subsection{Surface acoustic wave}

Along the  boundary between two ideal fluids any surface wave cannot propagate, because the  boundary conditions cannot be satisfied. But if the one of materials has one negative constitutive parameter boundary states can be excited~\cite{Ambati07}. This is a direct analogy  to surface plasmon states in plasma with $\epsilon<0$~\cite{Carloz05}.

Let us consider a wave traveling on the interface between a half-space occupied by pure host fluid and a half-space occupied by a fluid with embedded rods in it. Matching boundary impedances of evanescent waves in both composite and pure fluid, one can get:
\begin{equation}
\dfrac{\rho_e k^z_e - \rho_0 k^z_0}{\rho_e k^z_e + \rho_0 k^z_0} = 0,
\end{equation}
where  $k^z_0$ and $k^z_e$ are the projections of the wave-vector perpendicular to the interface in a pure fluid and metamaterial, respectively. Dispersion equation for surface wave has the following form:
\begin{equation}
 k_s=\sqrt{\dfrac{(\rho_0\rho_e)(\rho_0/B_e-\rho_e/B_0)}{\rho_0^2-\rho_e^2}}\omega.
\end{equation}
We show below, that the boundary states exist only if one of the constitutive parameters is negative.

\section{Results}
\subsection{Bulk wave}
In a model used for calculations the rods are considered to be two-layered cylinders. Shell of the cylinder is soft silicone resin\footnote{There are various types of silicone resins, therefore our parameters are approximate~\cite{Liu2000}}~($\rho = 1.3\cdot10^3kg/m^3$, $c_{11} = 5.2\cdot10^7N/m^2$, $c_{44} =3.25\cdot10^6N/m^2$, Loss: $600~dB/m$ at 1 MHz), core of the cylinder is  poly-crystal silver~\cite{Auldv1} ($\rho = 10.5\cdot 10^3 kg/m^3$, $c_{11} = 1.397\cdot10^{11}N/m^2$, $c_{44} =2.7\cdot10^{10}N/m^2$, Loss: $40~dB/m$ at 1 MHz). Outer radius of the rod is 3\,mm, radius of the core is 0.5\,mm, average distance between cylinders is taken 5\,mm. Host material is water~($\rho = 10^3kg/m^3$, $B = 2.25\cdot10^9N/m^2$).

\begin{figure}
\subfigure[]{\label{fig:s_eff}\includegraphics{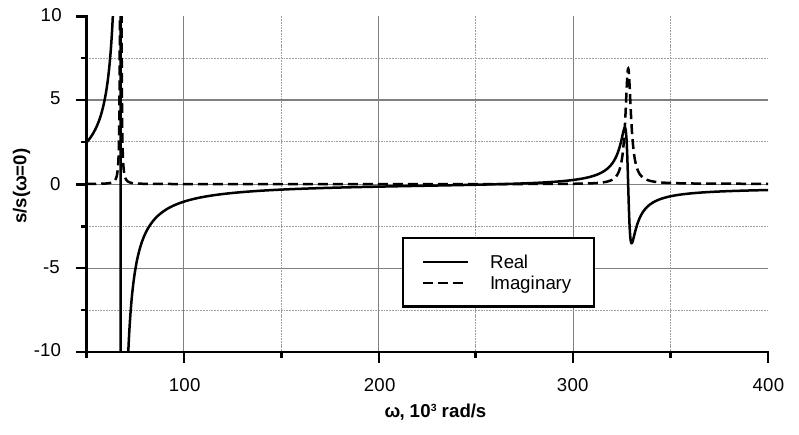}}
\subfigure[]{\label{fig:rho_eff}\includegraphics{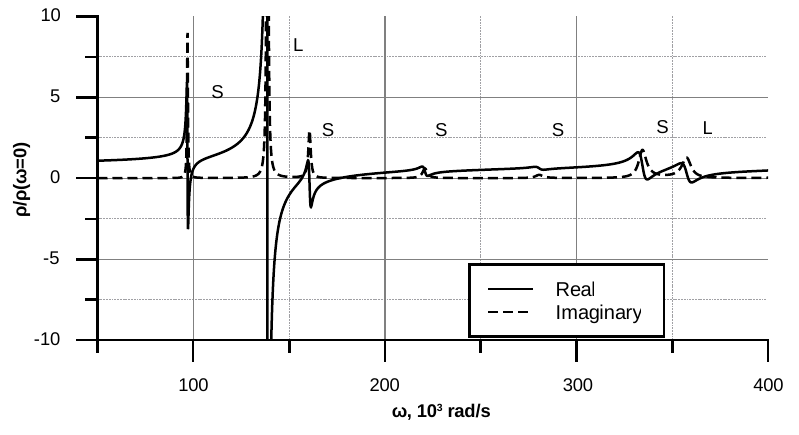}}
\subfigure[]{\label{fig:n_eff}\includegraphics{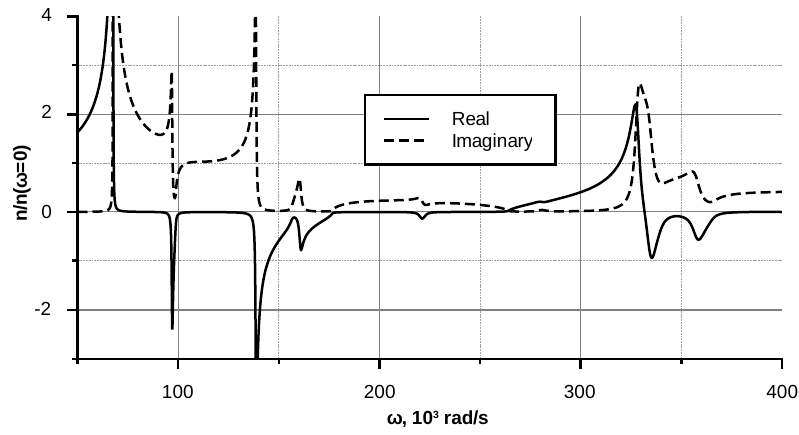}}
\subfigure[]{\label{fig:arg_n}\includegraphics{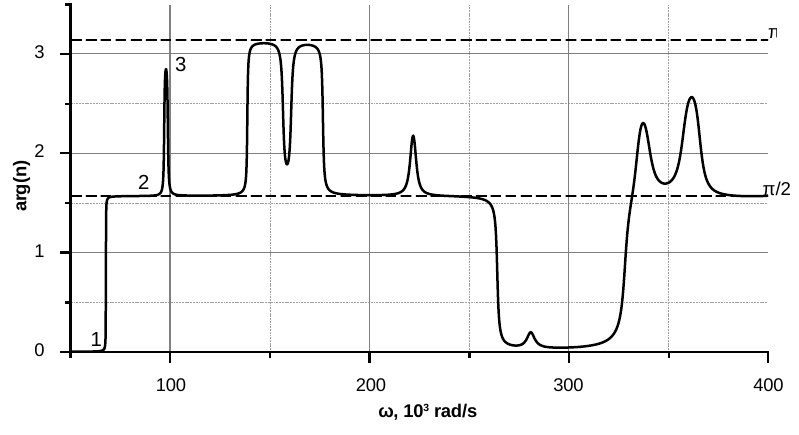}}
\caption{Frequency dependence of effective compliance~\subref{fig:s_eff}, density~\subref{fig:rho_eff}, refractive index~\subref{fig:n_eff} and the argument of refractive index~\subref{fig:arg_n} for composite, normalized on zero-frequency limit\label{fig:bulk_wave}}%
\end{figure}

Fig.~\ref{fig:bulk_wave} shows  results of calculations for bulk wave in the composite. At low frequencies effective constitutive parameters tend to be close to ``classic'' values for composites~\cite{Berryman1980}. But at frequencies close to resonances inside the inclusions the values of dynamical effective constitutive parameters are dramatically changed. In Fig.~\ref{fig:s_eff} two frequencies of monopolar resonances are seen and close to the resonances there are two bands where real part of compliance is negative (In this section we use compliance ($s$=$1/B$) instead of stiffness to obtain symmetry between constitutive parameters in~\eqref{fig:refind}). 

In Fig.~\ref{fig:rho_eff} frequency dependence of the dynamical effective density is shown. Due to the complex shear and longitudinal field structure inside  inclusion, there are two families of resonances associated with shear waves and longitudinal waves (S and L in the Figure, respectively). Thus dynamical effective density has a number of frequency ranges where real part is negative. The amplitude of resonances associated with shear components become greater because of inertia of the core.

In Figs.~\ref{fig:n_eff} and~\ref{fig:arg_n} dependence of the refractive index and its argument are plotted.   At the low frequency less than all resonance frequencies of inclusions  composite  exhibits properties of ``normal'' medium (region 1). At frequency greater that first monopolar resonance in the inclusions, the real part of $1/B$  become negative and the phase of wave vector is shifted to $\pi/2$ (region 2). In this range wave-number is close to be pure imaginary, thus no oscillations can be exited and no wave is able to propagate. In such way a band gap in the spectrum is formed. On frequency greater than dipole resonance, effective density become negative and phase is  shifted again for another $\pi/2$ and the wave-number becomes close to be real negative. Thus phase velocity of the wave points backwards and opposite to the Poynting vector. Also, dependence of refractive index shows, that in the frequency band where constitutive parameters have simultaneous significant negative parts, the imaginary part of refractive index does not dominate and the wave is still able to propagate despite of losses in inclusions.

\subsection{Surface acoustic wave}
\begin{figure}
  \centering
  \includegraphics{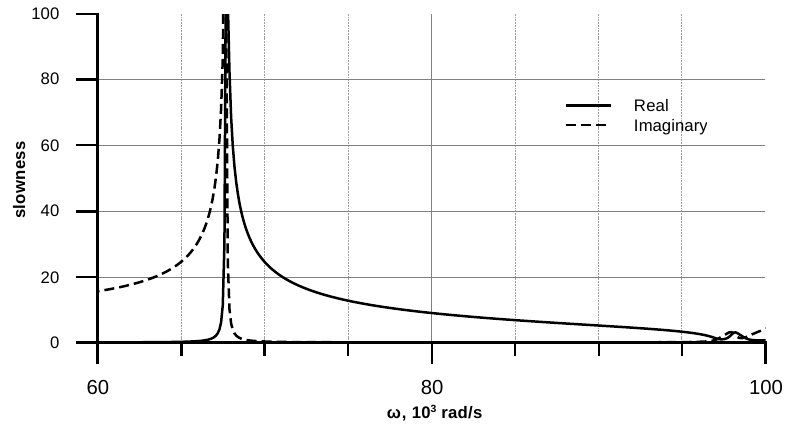}
  \caption{Slowness ($k_s/k_0$) of surface acoustic wave at the interface between metamaterial and ``normal'' material}
  \label{fig:saw}                
\end{figure}

Fig.~\ref{fig:saw} shows slowness ($k_s/k_0$) for surface wave to bulk wave in water. As it was predicted, wave-number does not have real part at low frequencies. Since the parameters of metamaterial are positive, the wave cannot propagate along the interface. But at frequency greater than resonance (compare to Fig.~\ref{fig:s_eff}) one of material parameters become negative and wave-number is real and the surface states appear. Surface wave is much slower than bulk wave, thus the surface states are bounded to the interface.

\begin{figure}
  \centering
  \includegraphics{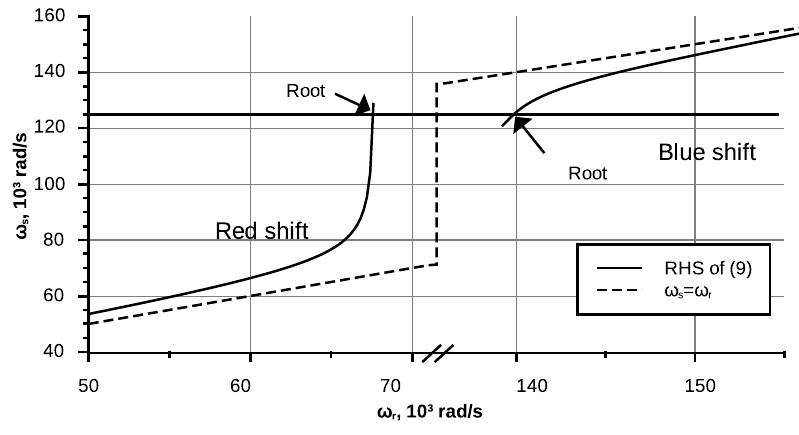}
  \caption{Graphical solution for equation~\eqref{eq:doppler} in metamaterial}
  \label{fig:doppler} 
\end{figure}

\subsection{Complex Doppler effect}

It was shown that in electromagnetic media with simultaneous negative $\epsilon$ and $\mu$ Doppler effect should be reversed~\cite{Veselago68}. Similar effect can exist for  acoustic metamaterials. Let us consider a source travels inside a composite with velocity $V$. It  transmits signal with an angular frequency $\omega_s$. We can assume that source is moving in one line with receiver. To find frequency on the receiver~($\omega$) we should solve the following equation:
\begin{equation}
   \omega_s = \omega - V\cdot k(\omega)
\label{eq:doppler}
\end{equation}
For dispersive medium like metamaterial solution for equation~\eqref{eq:doppler} is not trivial~\cite{Berger76}. 

Graphical solution for equation~\eqref{eq:doppler} is plotted in Fig.~\ref{fig:doppler}. We assume that dumping is small and and wave-number is real. Angular frequency of source is taken $1.22\cdot10^5$ rad/s and speed is 15 m/s. The source moves away from receiver.  For the pure host material the result is predictable, the frequency on receiver is down-shifted. In contrast, for composite equation~\eqref{eq:doppler} has two roots and the two modes with down-shifted and up-shifted frequencies are present. Generation of two modes is possible, because modes have different phase velocities and one of the waves if forward and another is backward~\cite{Berger76}.

\section{Conclusions}
In this paper effective dynamic density and stiffness of composite consisted of water with embedded solid two-layered rods in terms of coherent potential approximation are calculated. It is shown, that there are some frequency ranges in which dynamic constitutive parameters are simultaneously negative and the propagating wave becomes backward.

Dispersion of the surface acoustic wave at the interface between metamaterial and ``normal'' material is calculated. It is shown that there are frequency ranges in which surface states are bounded to the interface. 

Doppler effect in composite is considered. It is shown, that in metamaterial Doppler shift is reversed and several modes could be excited.

\begin{acknowledgement}
The work is supported by RFBR (Grants \#11-02-01132-a, \#08-02-00785-a, \#09-02-12433-ofi-m,  \#09-07-13531-ofi-ts), GPAD RAS (Program ``Coherent Fields and Signals'') and Ministry of Education and Science of RF (agreement \#$\Pi556$)
\end{acknowledgement}


\end{document}